\pdfoutput=1

\documentclass[11pt]{article}

\usepackage[preprint]{acl}

\usepackage{times}
\usepackage{latexsym}

\usepackage[T1]{fontenc}

\usepackage[utf8]{inputenc}

\usepackage{microtype}

\usepackage{inconsolata}

\usepackage{graphicx}
\usepackage{xspace}
\usepackage{amsmath}
\usepackage{wrapfig}
\usepackage[ruled,vlined]{algorithm2e}
\usepackage{listings}
\usepackage{enumitem}
\usepackage[breakable]{tcolorbox}
\usepackage{booktabs}
\usepackage{multirow}

\newcommand{\leetcode}[0]{LeetCode\xspace}

\newcommand{{\method}}[0]{\textsc{TestEval}\xspace}

\title{{\method}: Benchmarking Large Language Models for Test Case Generation}

\author{
  \textbf{Wenhan Wang\textsuperscript{1}$^*$},
  \textbf{Chenyuan Yang\textsuperscript{3}$^*$},
  \textbf{Zhijie Wang\textsuperscript{1}$^*$},
  \textbf{Yuheng Huang\textsuperscript{2}},
\\
  \textbf{Zhaoyang Chu\textsuperscript{4}},
  \textbf{Da Song\textsuperscript{1}},
  \textbf{Lingming Zhang\textsuperscript{3}},
  \textbf{An Ran Chen \textsuperscript{1}},
  \textbf{Lei Ma \textsuperscript{2,1}}
\\
  \textsuperscript{1}University of Alberta,
  \textsuperscript{2}The University of Tokyo,
  \textsuperscript{3}University of Illinois Urbana-Champaign, \\
  \textsuperscript{4}Huazhong University of Science and Technology
\\
\texttt{wenhan12@ualberta.ca} \ \  \texttt{cy54@illinois.edu} \ \ \texttt{zhijie.wang@ualberta.ca} \\ \texttt{yuhenghuang42@g.ecc.u-tokyo.ac.jp} \ \ \texttt{chuzhaoyang@hust.edu.cn} \ \ \texttt{dsong4@ualberta.ca} \\ \texttt{lingming@illinois.edu} \ \  \texttt{anran6@ualberta.ca} \ \ \texttt{ma.lei@acm.org}
}

\begin{document}
\maketitle
\def\thefootnote{*}\footnotetext{These authors contributed equally to this work.}\def\thefootnote{\arabic{footnote}}

\begin{abstract}
For program languages, testing plays a crucial role in the software development cycle, enabling the detection of bugs, vulnerabilities, and other undesirable behaviors. To perform software testing, testers need to write code snippets that execute the program under test. Recently, researchers have recognized the potential of large language models (LLMs) in software testing. However, there remains a lack of fair comparisons between different LLMs in terms of test case generation capabilities.

  In this paper, we propose {\method}, a novel benchmark for test case generation with LLMs. We collect 210 Python programs from an online programming platform, \leetcode, and design three different tasks: overall coverage, targeted line/branch coverage, and targeted path coverage. We further evaluate 17 popular LLMs, including both commercial and open-source ones, on {\method}. We find that generating test cases to cover specific program lines/branches/paths is still challenging for current LLMs, indicating a lack of ability to comprehend program logic and execution paths. We have open-sourced our dataset and benchmark pipelines at \url{https://github.com/LLM4SoftwareTesting/TestEval}.
\end{abstract}

\section{Introduction}

Software testing is a crucial aspect of software development, allowing developers to identify potential bugs and check if the program behavior meets expectations. A key task in software testing is test case generation, which involves creating test inputs to cover different statements and branches in the program under test. Previous research indicates that test case generation can be extremely time-consuming, taking up over 15\% of the time spent in software development~\cite{daka2014survey}.

Therefore, automated test case generation has been a long-standing challenge in software engineering research. Various methods have been developed to address this issue, including symbolic execution testing~\cite{chipounov2011s2e, cadar2011symbolic}, search-based testing~\cite{fraser2011evosuite, baresi2010testful, fraser2010mutation}, and deep learning-based approaches \cite{tufano2020unit}. 
Recently, researchers have been exploring the potential of using LLMs to generate unit test cases~\cite{lemieux2023codamosa, xie2023chatunitest, yuan2023no}. However, despite the rapid development of LLM-based test case generation, there is still a lack of public benchmarks to evaluate different LLMs' capabilities in this area. 
Hence, there is a need for a comprehensive analysis to determine whether current LLMs can (1) generate diverse test cases to achieve high coverage on a program under test, (2) generate test cases to cover a specific line or branch, and (3) generate test cases to cover a specific execution path by following the tester's intent.

To bridge this gap, we present a new benchmark, {\method}, which focuses on evaluating LLMs' test case generation capabilities. The {\method} dataset consists of 210 Python programs from the online coding platform \leetcode. We design three tasks to address the aforementioned challenges: (1) overall coverage, (2) targeted line/branch coverage, and (3) targeted path coverage.

Notably, unlike popular code generation benchmarks \cite{chen2021evaluating, austin2021program} or software testing datasets \cite{just2014defects4j, lemieux2023codamosa}, the tasks in our {\method} benchmark require LLMs to reason about intricate program execution behaviors. To generate inputs that invoke specific branches or paths in the program under test, the LLM must have a comprehensive understanding of how to satisfy certain branch conditions during execution. Furthermore, our tasks emphasize program logic analysis rather than merely simulating numerical operations, as seen in benchmarks designed for predicting a program's input/output~\cite{gu2024cruxeval}.

\begin{figure}[t]
  \centering
  \vspace{-10pt}
  \includegraphics[width=\columnwidth]{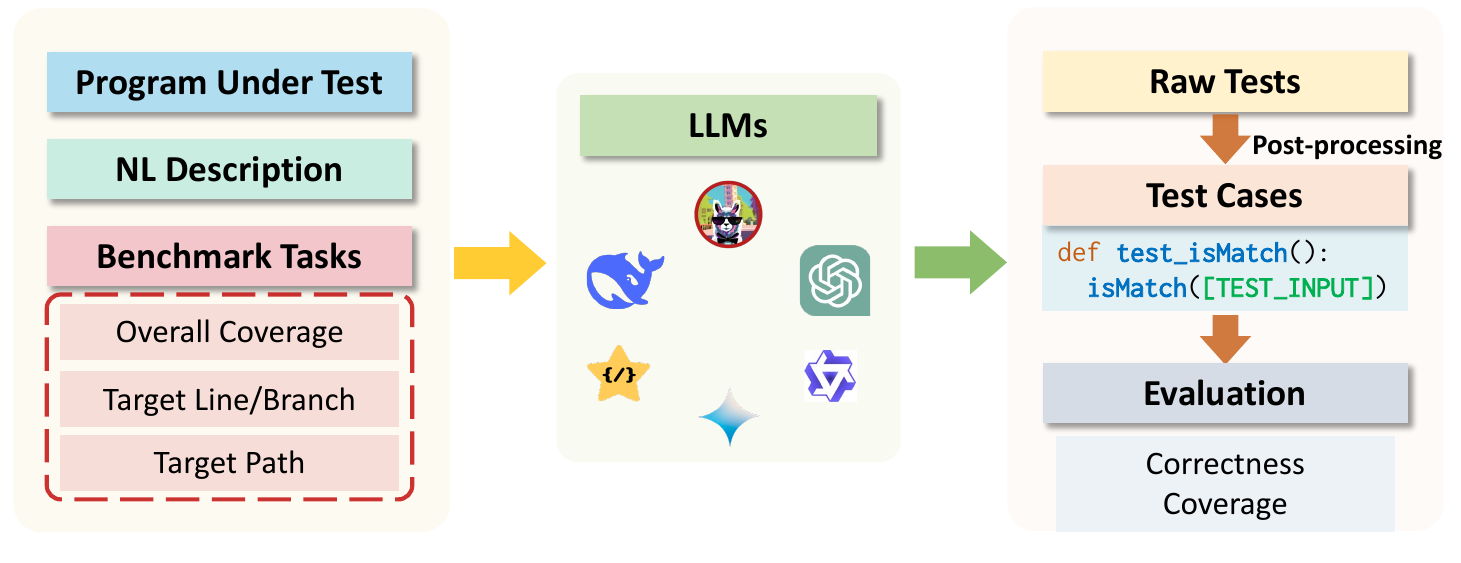}
  \vspace{-10pt}
  \caption{The pipeline for running and evaluating LLMs for test case generation on {\method}.}
  \vspace{-10pt}
  \label{fig:workflow}
\end{figure}

We perform extensive experiments on {\method} with both commercial and open-source LLMs. Our results indicate that while state-of-the-art LLMs can generate executable and diverse test cases, they struggle to identify which specific statements or branches need to be covered.  For example, in targeted line coverage, 12 out of 16 LLMs' performances are not significantly improved (improvements $\leq 5\%$) compared to the results when the target line information is even \textbf{not} given.
Quantitative results show that commercial LLMs, such as GPT-4, generally outperform open-source LLMs in both overall coverage and targeted line/branch/path coverage. These findings suggest that future work on test case generation should focus on developing advanced LLM-based reasoning frameworks to enhance the understanding of program behaviors during testing.

Our work makes the following contributions:

\begin{itemize}[leftmargin=*]
    \item \textbf{Benchmark.} We propose {\method}, a benchmark focused on evaluating LLMs' capabilities in generating test cases for a given program under test, encompassing three different tasks.

    \item \textbf{Evaluation.} We design new metrics to measure the LLM's test generation performance and conduct experiments with 17 popular LLMs.

    \item \textbf{Analysis.} We perform a systematic analysis of LLMs' performance on {\method} and discuss the challenges and opportunities in test case generation using LLMs.

\end{itemize}

\section{Approach}

In this section, we first introduce the tasks included in our benchmark (\S~\ref{subsec:task}). Following that, we provide an overview of the dataset used (\S~\ref{subsec:dataset}).

\subsection{Task Description}
\label{subsec:task}

Figure~\ref{fig:workflow} shows the workflow of {\method}. We propose three distinct tasks in our benchmark: (1) overall coverage, (2) targeted line and branch coverage, and (3) targeted path coverage. For each task, we prompt an LLM to generate test cases for a specified program based on the task description in natural language. Specifically, in each query round, we prompt the LLM to generate a testing function containing a single test case (see Appendix~\ref{appendix:template} for the complete prompt templates). Then, we filter out any non-code content that may have been generated outside the testing function, retaining only the first test case generated in each query round to ensure a fair comparison across different LLMs.

After generation, all test cases must undergo a correctness check, which consists of \textit{syntactic correctness}, \textit{execution correctness}, and \textit{assertion correctness}. Syntactic correctness determines if the generated test case is free of syntax errors, while execution correctness evaluates if the test case can be executed successfully without any runtime errors. Assertion correctness evaluates whether the generated test case contains correct test assertions. Regarding execution correctness, we do not consider incorrect test assertion statements to be failed cases, since test cases with assertion errors can still cover the program under test. Finally, we evaluate coverage metrics on test cases that pass the correctness check. We now illustrate our three benchmark tasks in detail.

\begin{algorithm}[h]
\scriptsize
\label{alg:cov@k}
\SetAlgoLined 
\caption{Computing the average line/branch $cov@k$ given a set of programs}
\KwIn{A set of programs under test $\mathcal{P}=\left\{ p_{1}, p_{2},...\right\}$, $k$}
\KwOut{The average $cov@k$ for all programs: $cov@k_{all}$}
$cov@k$ = []\;
\For{$p_i$~in~$\mathcal{P}$}{
    Generate $N$ test cases $T_{i}=\left\{ t_{i1}, t_{i2},..., t_{iN}\right\}$\;
    Retain $M$ executable test cases $T_{i}=\left\{ t_{i1}, t_{i2},..., t_{iM}\right\}$\;
    \eIf{$T_{i}=\emptyset$}{
        $cov@k$.append($0$)\;
    }
    {
    Randomly split $T_{i}$ into $ \textrm{max}(\lfloor M/k \rfloor, 1)$ groups, each group $T_{ij}$ with $k$ test cases\;
    $cov_{i}$ = []\;
    \For{$T_{ij}$~in~$\left\{ T_{i1}, T_{i2},..., T_{i\lfloor M/k \rfloor}\right\}$}
    {
        Compute line/branch coverage $cov_{T_{ij}}$\;
        $cov_{i}$.append($cov_{T_{ij}}$)\;
    }
    $cov@k$.append(\textrm{avg}($cov_{i}$)) \\
    }
}
$cov@k_{all}~\gets~\textrm{avg}(cov@k)$\;
\Return{$cov@k_{all}$}
\end{algorithm}

\textbf{Overall coverage}. In this task, we query each LLM for $N$ rounds given a program under test. During the $i$th ($1<i\leq N$) round, we prompt the LLM to generate a test method different from the $j\ (1\leq j<i)$th rounds. After all rounds of query, we obtain $N$ test cases for each program under test.
The overall coverage for a program is computed by the proportion of lines/branches in the program that have been covered by at least one test case.

We further propose a new metric, $cov@k$, to measure the diversity of LLM's generated test cases for a given program. Intuitively, $cov@k$ measures the line/branch coverage with a subset of the generated test cases with a size of $k$ ($k < M$). To achieve this, we randomly split $M$ executable test cases into $max(\lfloor M/k \rfloor, 1)$ subsets. Then for each of these subsets, we calculate its overall line/branch coverage. In our experiments, we choose $k$ as 1, 2, and 5. When $k$ increases, the improvements of $cov@k$ can measure the diversity of the LLM's generated test cases. We summarize the calculation of the average line/branch $cov@k$ for a set of programs under test in Algorithm~1.

\begin{figure}[t]
  \begin{center}
    \includegraphics[width=\columnwidth]{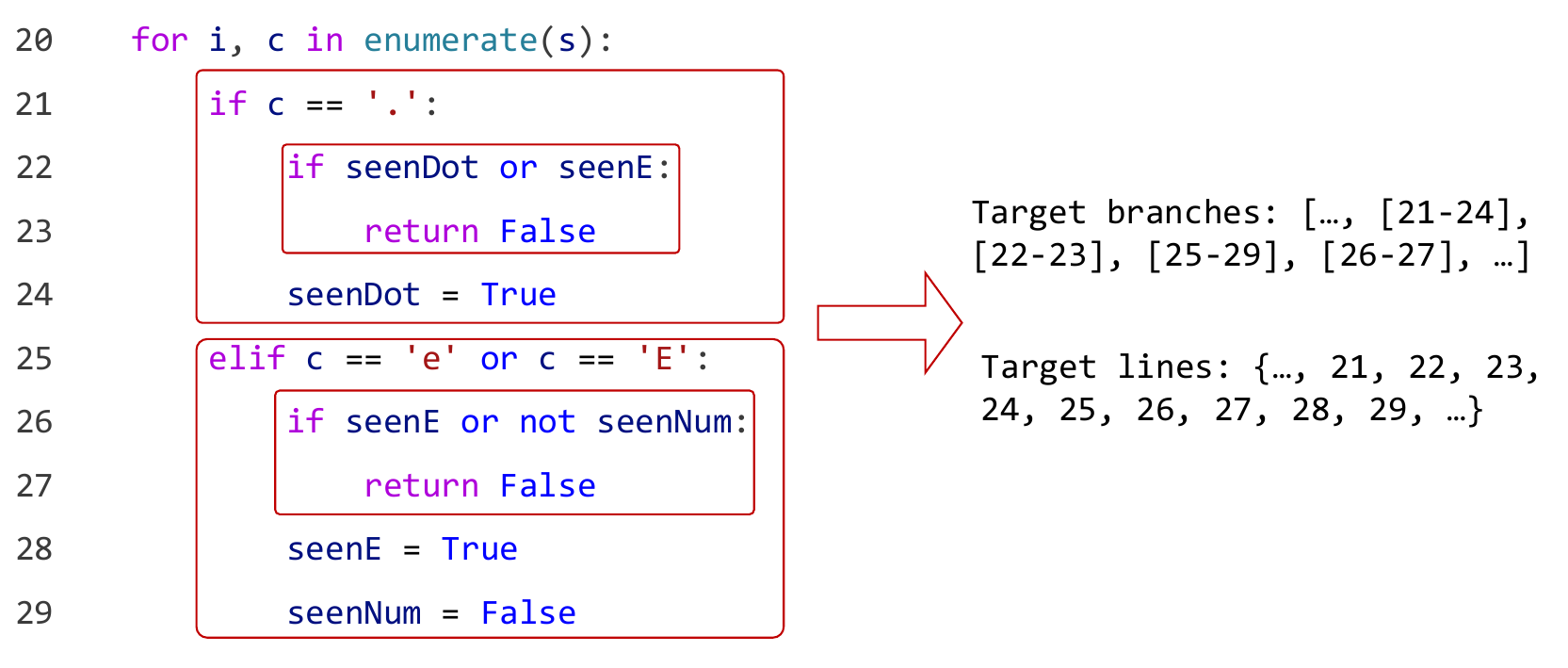}
  \end{center}
  \caption{An example for selecting targeted lines and branches from programs under test.}
  \label{fig:linebranch}
\end{figure}

\textbf{Targeted line and branch coverage.} Different from overall coverage, \textit{targeted line and branch coverage} requires the LLM to generate test cases that could cover a specific branch, or a line inside this branch. This simulates the scenario in which a human tester is asked to craft test cases to cover a specific part of the program. Figure~\ref{fig:linebranch} shows an example of targeted branches and lines in a given program. To measure the targeted line/path coverage, we prompt the LLMs by including the line number(s) in the instruction (see prompt templates in Appendix~\ref{appendix:template}). For each targeted line/branch, an LLM is prompted to generate one test case. 

\begin{figure*}[h]
  \centering
    \includegraphics[width=0.8\textwidth]{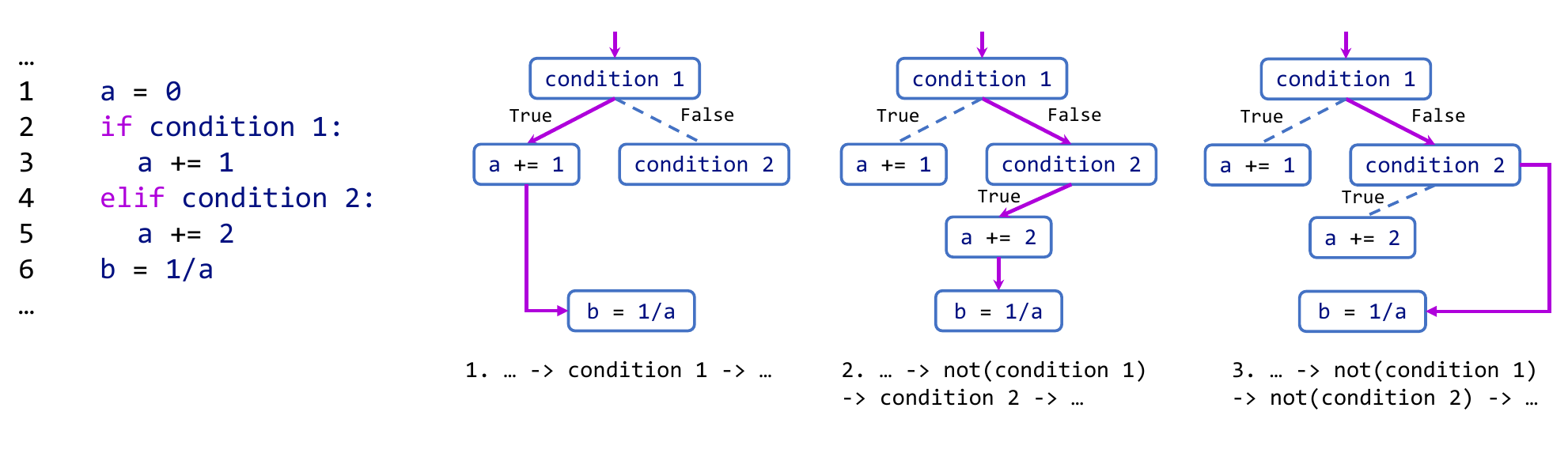}
  \caption{A motivating example showing the importance of path coverage (left), and examples of execution paths extracted from this program (right).} 
  \label{fig:path}
\end{figure*}

\textbf{Targeted path coverage.} In real-world software development, testers sometimes need to craft test cases to cover a specific execution path that includes multiple branches. We refer to this task as the \textit{target path coverage}. We show an example program in Figure~\ref{fig:path} to demonstrate the importance of the target path coverage.

In Line 6, a bug (divided by zero) will occur only if branches ``condition 1'' and ``condition 2'' are both \textbf{not} executed. In this case, only covering the two conditional branches is not sufficient. By contrast, if we can cover all three paths (Figure~\ref{fig:path}), we can successfully detect the ``divided by zero'' bug. To obtain the target path coverage, we prompt an LLM by including a specific execution path (see Appendix~\ref{appendix:template} for the prompt template). For each path, an LLM is queried to generate one test case.

We further propose two metrics to evaluate the performance of target path coverage. First, for a given target path, we measure whether the generated test case covers the target path completely. Second, we measure the similarity between the given target path $\text{PATH}_{tgt}$ and the execution path of the LLM's generated test case $\text{PATH}_{gen}$ by Eq.~\ref{eq:path_sim}:

\begin{equation}
\scriptsize
\label{eq:path_sim}
    sim(\text{PATH}_{gen}, \text{PATH}_{tgt})= \frac{lcs(\text{PATH}_{gen}, \text{PATH}_{tgt})}{len(\text{PATH}_{tgt})}
\end{equation}
where $lcs()$ calculates the length of the longest contiguous common sub-sequence between two paths and $len()$ calculates the length of a path.

\subsection{Benchmark Dataset}
\label{subsec:dataset}

\textbf{Data collection.} To construct our benchmark dataset, we first collect solution programs of LeetCode, an online platform for evaluating a programmer's coding performance. We choose LeetCode as our data source since it has a clear task description and input constraint for each programming task. We first select all publicly available tasks up to Apr. 2024. Then, we collect its Python solution for each task from a GitHub repository~\footnote{https://github.com/walkccc/LeetCode. The repository is under MIT license.}. At this stage, we collect 3,123 programs under test.

The main goal of our benchmark is to evaluate LLMs' capability to generate test cases that cover specific statements/branches. Therefore, we filter out programs that are too simple (e.g., programs that only have one branch) according to the cyclomatic complexity~\cite{mccabe1976complexity}. Given the control flow graph of a program, the cyclomatic complexity $V$ of this program is measured by: $V=e-n+p$, where $e$ is the number of edges in the graph, $n$ is the number of nodes, and $p$ is the number of connected components. The cyclomatic complexity is positively correlated with the number of branches/loops in a program. In this work, we consider programs with the cyclomatic complexity $\geq$ 10. This filters down the sample size from 3,123 to 216. We further check these 216 problems and remove similar problems with identical solutions.
Finally, we collect 210 Python programs for our benchmark, consisting of 9 easy problems, 100 medium problems, and 101 hard problems according to \leetcode's difficulty label. Each program under test is also paired with its task description in natural language. Note that most programs already have test cases in their task descriptions. We remove these cases to prevent LLMs from directly copying these test cases. 

For each program, we perform the following pre-processing steps:

\begin{itemize}
    \item We add all necessary \texttt{import} statements for the packages required by the Python solution.
    \item Python programmers often split long statements into multiple physical lines. For all statements split into multiple lines, we rewrite them in a single line. This ensures each statement only corresponds to one line when measuring line coverage.
    \item We reformat the in-line conditional statements (e.g., the ternary conditional operator) into multi-line blocks. This ensures that each line of the program is one statement that belongs to one specific branch.
    \item We remove all natural language comments.
\end{itemize}

\textbf{Targeted line/branch/path identification.} To obtain targeted lines/branches, we first extract all \textit{conditional} branches of a given program based on its abstract syntax tree (AST). Since loop branches (i.e., \texttt{for}/\texttt{while} loops) are usually easy to cover, we only consider conditional branches in our task. Specifically, we extract all \texttt{if}, \texttt{elif}, and \texttt{else} branches. We refer to these branches as our targeted branches. Then, we consider all statements within these targeted branches as our targeted lines (see Figure~\ref{fig:linebranch} for the example). Overall, we identified 983 target branches in 210 programs under test (4.7 target branches per program on average). The total number of target lines in 210 programs is 1,312 (6.2 target lines per program on average). The detailed algorithm for extracting target lines/branches can be found in Appendix~\ref{appendix:branch}.

\begin{figure*}[h]
  \centering
    \includegraphics[width=0.8\textwidth]{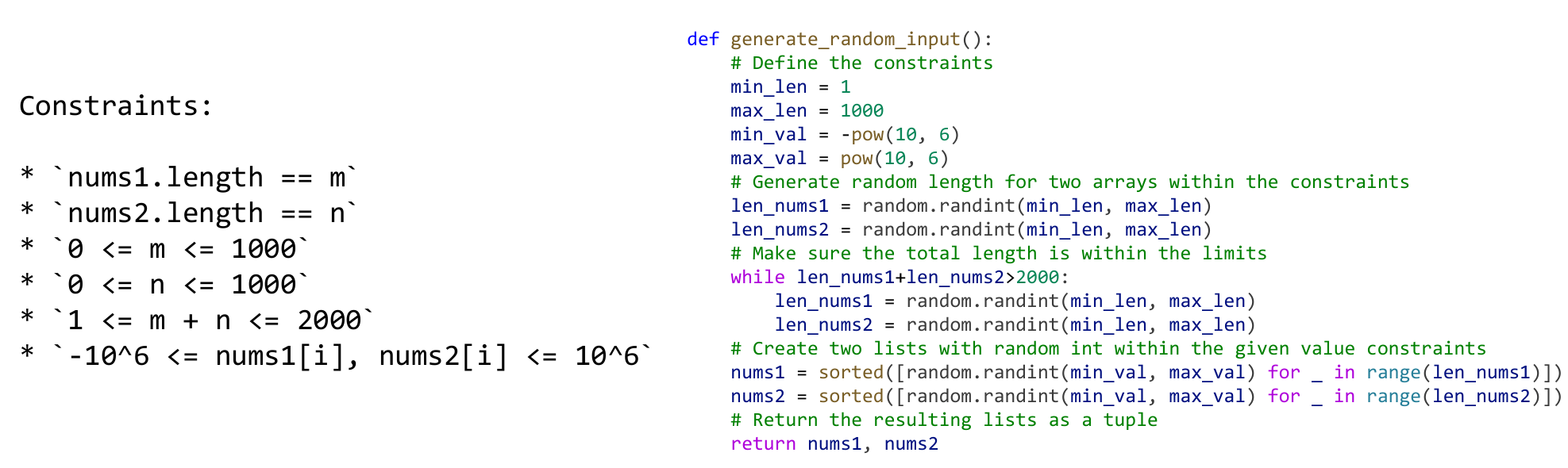}
  \caption{The input constraints for a LeetCode problem (left) and its random input generator for \method (right).} \label{fig:generator}
\end{figure*}

In a given program under test, certain branches could be hard to cover without carefully crafting the test cases. Therefore, we label each targeted branch in a program as easy, medium, or hard according to the average coverage after executing 100 randomly generated inputs. 
For each problem in \method, we construct a \emph{random input generator} to assess the difficulties of covering a specific branch. Each generator is a Python program that \emph{uniformly} samples a valid test input for the LeetCode problem according to its constraint description.
We leverage GPT-4 to generate these generators from the constraints in \leetcode problem descriptions. Then, we perform manual inspection and correction to ensure they adhere to the problem's constraints. An example of an input generator is shown in Figure~\ref{fig:generator}.
These generators are then used to sample 100 executable test cases for each problem. Branch difficulty is determined by the frequency at which a branch is covered across theese 100 sampled test inputs.
We categorize branches as follows: easy (covered by $[99\%,100\%]$ of test cases), medium (covered by $[40\%,99\%)$ of test cases), and hard (covered by $[0,40\%)$ of test cases). 
This partitioning ensures that easy branches do not significantly outnumber other categories, and promotes a balanced distribution between medium and hard branches. The number of easy, medium, and hard target branches are 498, 225, and 260.

For the targeted path coverage task, as the number of execution paths in a program can be enormous or even undecidable, it is impossible to collect all execution paths. Instead, we collect the target execution paths from the example test cases given by \leetcode problem descriptions. For each example test case, we execute it and record its execution path using all the condition/loop branches it executed. The complete execution path would be too long and difficult for LLMs to understand, so we perform clipping after obtaining full paths. For each execution path, we randomly sample two sub-paths with lengths of 5 consecutive branches taken. We further remove duplicated sampled paths, resulting in an average of 4.1 target paths per problem.

\section{Evaluation}


\subsection{Experiment Setup}
\label{subsec:setup}

We evaluate 17 popular instruction-following LLMs, including both commercial and open-source ones. The parameter sizes of open-source models range from 1.3B to 34B. The temperature is set to \texttt{0} or \texttt{1e-5} (for models on Huggingface that do not support \texttt{temperature=0}) to ensure that the evaluation results can be reproduced. All experiments on open-source LLMs are run on two NVIDIA A6000 GPUs. We set the length limit of outputs to 256 tokens. We use the \texttt{pytest-cov}~\cite{pytest-cov} to measure the code coverage.

\begin{table*}[h]
  \caption{Result on the overall coverage task. The results in parenthesis are the improvements over $cov@1$.}
  \label{overall-coverage}
  \centering
  \scriptsize
  \setlength{\tabcolsep}{1.8pt}
  \scalebox{0.9}{
  \begin{tabular}{llcccccccccp{0.5cm}lp{0.5cm}lccp{0.5cm}lp{0.5cm}l}
    \toprule
    \multirow{2}{*}{Model} & \multirow{2}{*}{Size} & & \multicolumn{3}{c}{Correctness} & & \multicolumn{2}{c}{Overall coverage} & & \multicolumn{5}{c}{Line $cov@k$} & & \multicolumn{5}{c}{Branch $cov@k$}                 \\
    \cmidrule(lr){4-6} \cmidrule{8-9} \cmidrule(lr){11-15} \cmidrule(l){17-21}
    & & & syntax & execution & assertion & & line & branch & & $k=1$ & \multicolumn{2}{c}{$k=2$} & \multicolumn{2}{c}{$k=5$} & & $k=1$ & \multicolumn{2}{c}{$k=2$} & \multicolumn{2}{c}{$k=5$}\\
    \midrule
    GPT-3.5-turbo & N/A &  & \textbf{100} & 97.43 & 40.40 & & 96.27 & 93.65 & & 88.35 & 90.02 & (1.67) & 92.14 & (3.79) & & 81.87 & 84.32 & (2.45) & 87.55 & (5.68)\\
    GPT-4  & N/A &  & \textbf{100} & 92.33 & 54.16 & & 94.94 & 92.81 & & 85.65 & 87.77 & (2.12) & 90.04 & (4.39) & & 78.89 & 81.93 & (3.04) & 85.39 & (6.50)\\
    GPT-4-turbo & N/A &   & \textbf{100} & 94.79 & 56.24 & & 96.08 & 94.81 & & 85.46 & 87.87 & (2.41) & 90.81 & (5.35) & & 78.62 & 82.06 & (3.44)	& 86.64 & (8.02)\\
    GPT-4o & N/A & & 99.59 & \textbf{98.30} & \textbf{52.99} & & 98.65 & 97.16 & & \textbf{90.23} & \textbf{92.16} & (1.93) & \textbf{94.33} & (4.10) & & \textbf{84.05} & \textbf{86.89} & (2.84) & \textbf{90.31} & (6.26)\\
    GPT-4o-mini & N/A & & \textbf{100} & \textbf{99.92} & 43.86 & & \textbf{98.76} & \textbf{97.58} & & 88.06 & 90.33 & (2.27) & 93.51 & (5.45) & & 81.64 & 85.03 & (3.39) & 89.60 & (7.96)\\
    \midrule
    Gemini-1.0-pro & N/A & & 93.05 & 71.93 & 35.31 & & 93.01 & 90.66 & & 84.48 & 86.60 & (2.12) & 88.47 & (3.99) & & 78.35 & 81.29 & (2.94) & 84.11 & (5.76)\\
    \midrule
    \multirow{3}{*}{CodeLlama} & 7b & & 99.52 & 73.86 & 31.07 & & 86.09 & 81.56 & & 79.46 & 80.72 & (1.26) & 82.04 & (2.58) & & 72.28 & 73.96 & (1.68) & 75.90 & (3.62)\\
    & 13b & & 67.55 & 50.40 & 25.28 & & 85.66 & 80.55 & & 80.49 & 82.26 & (1.77) & 83.44 & (2.95) & & 73.21 & 75.54 & (2.33) & 77.13 & (3.92)\\
    & 34b & & 66.33 & 46.86 & 40.32 & & 87.96 & 83.74 & & 78.83 & 81.25 & (2.42) & 83.71 & (4.88) & & 71.37 & 74.50 & (3.13) & 77.80 & (6.43)\\
    \midrule
    Llama3 & 8b & & 99.25& 82.24 & 44.61 & & 90.98 & 89.02 & & 77.40 & 80.08 & (2.68) & 84.42 & (7.02) & & 69.47 & 73.37 & (3.90) & 79.22 & (9.75)\\
    \midrule
    Llama3.1 & 8b & & 98.69 & 94.69 & 50.00 & & 88.94 & 85.79 & & 74.42 & 77.49 & (3.07) &  82.07 & (7.65) & & 65.65 & 69.92 & (4.27) & 76.16 & (10.51)\\
    \midrule
    Gemma & 7b & & 98.98 & 64.64 & 35.30 & & 93.16 & 91.46 & & 76.23 & 80.54 & (4.31) & 85.90 & (9.67) & & 67.15 & 72.94 & (5.79) & 80.29 & (13.14)\\
    \midrule
    Starcoder-2-Instruct & 15b & & 97.07 & 94.07 & 54.11 & & 89.84 & 84.41 & & 88.03 & 88.22 & (0.19) & 88.50 & (0.47) & & 81.80 & 82.09 & (0.29) & 82.50 & (0.70)\\
    \midrule
    \multirow{3}{*}{DeepSeek-coder} & 1.3b & & 96.05 & 82.48 & 38.66 & & 81.22 & 75.99 & & 75.89 & 76.50 & (0.61) & 77.09 & (1.20) & & 69.06 & 69.90 & (0.84) & 70.70 & (1.64)\\
    & 6.7b & & 97.42 & 82.43 & 40.43 & & 93.48 & 91.61 & & 82.40 & 84.74 & (2.34) & 87.97 & (5.57) & & 75.29 & 78.73 & (3.44) & 83.46 & (8.17) \\
    & 33b & & 99.21 & 83.57 & 50.75 & & 94.86 & 91.92 & & 85.47 & 87.38 & (1.91) & 90.30 & (4.83) & & 78.49 & 81.23 & (2.74) & 85.12 & (6.63)\\
    \midrule
    CodeQwen & 7b & & \textbf{100} & 84.26 & 46.36 & & 90.73 & 86.90 & & 84.53 & 85.33 & (0.80) & 86.71 & (2.18) & & 77.66 & 78.94 & (1.28) & 80.95 & (3.29) \\
    \bottomrule
  \end{tabular}
  }
\end{table*}

\subsection{Overall Coverage}
In this experiment, we query every model 20 rounds ($N=20$) to generate test cases (one test case per round) for each program under test. Table~\ref{overall-coverage} shows the evaluation results on the overall coverage task.

Regarding correctness metrics, we observe that most models can achieve high syntactical and acceptable execution correctness, but all models have much lower assertion correctness.
For test cases that do not pass the execution correctness check, we perform a preliminary study in the Appendix \ref{sec:error-exec}.
Regarding the coverage performance, 
most of the LLMs are able to generate test cases that cover over 80\% lines/branches per program under test. Notably, the latest GPT-4o achieves the best overall line (98.65\%) and branch (97.16\%) coverage. We also notice that the open-source model, DeepSeek-coder-33b, outperforms the commercial LLM, Gemini-1.0-pro, on both overall line and branch coverage.

We further use $cov@k$ to measure the diversity of each LLM's generated test cases. Similar to the overall coverage results, GPT-4o has the best line and branch $cov@1$, demonstrating its ability to craft complex test cases that are able to cover most of the program branches within a single attempt. We also find all LLMs have a higher $cov@2$ and $cov@5$ compared with $cov@1$. This indicates that the LLMs are able to generate different test cases. Gemma-7b shows the most significant improvements in the line (+9.67\%) and branch (+13.14\%) $cov@5$ compared with its line and branch $cov@1$. We also notice that Starcoder-2-Instruct has the least improvement on $cov@5$ compared with $cov@1$ (+0.47\% and +0.70\% for line and branch coverage, respectively). By manually checking the test cases generated by Starcoder-2-Instruct, we find that it frequently repeats previously generated cases despite being instructed to generate different ones.

\begin{table*}[h]
  \caption{Results for targeted line coverage. Results in parenthesis are the improvements over baselines.}
  \label{line-coverage}
  \centering
  \scriptsize
  \scalebox{0.8}{
  \begin{tabular}{llcccclcccc}
    \toprule
    \multirow{2}{*}{Model} & \multirow{2}{*}{Size}  & \multicolumn{5}{c}{Targeted line} & & \multicolumn{3}{c}{Baseline: no targeted line} \\
    \cmidrule{3-7} \cmidrule{9-11}
    & &  syntax & execution & assertion & \multicolumn{2}{c}{cov. Recall} & & Syntax & execution & cov. Recall\\
    \midrule
    GPT-3.5-turbo & N/A  &  99.40 & 95.67 & 41.93 & 67.76 & (-1.27) & & \textbf{100} & \textbf{100} & 69.03\\
    GPT-4 & N/A  &  \textbf{100} & 98.81 & 61.22 & 78.20 & (10.14) & & \textbf{100} & 99.52 & 68.06\\
    GPT-4-turbo & N/A  &   99.20 & 98.73 & 67.29 & 80.52 & \textbf{(11.64)} & & \textbf{100} & \textbf{100}	& 68.88\\
    GPT-4o & N/A &  99.63 & 98.96 & \textbf{67.52} & \textbf{80.97} & (9.48) & & \textbf{100} & \textbf{100} & \textbf{71.49}\\
    GPT-4o-mini & N/A & \textbf{100} & \textbf{99.92} & 56.02 & 76.94 & (8.73) & & \textbf{100} & \textbf{100} & 68.21\\
    \midrule
    Gemini-1.0-pro & N/A  & 100 & 96.04 & 53.37 & 70.75 & (4.93) & & \textbf{100} & 95.71 & 65.82\\
    \midrule
    \multirow{3}{*}{CodeLlama} & 7b  & 99.85 & 90.97 & 34.04 & 58.13 & (0.89)	& & 99.52	& 93.81	& 57.24\\
    & 13b  & 99.63 & 85.22 & 48.42 & 54.63 & (-4.03)	& & 99.05	& 94.76	& 58.66\\
    & 34b  & 98.66 & 90.60 & 44.34 & 59.48 & (-0.29) & & \textbf{100} & 96.19 & 59.77\\
    \midrule
    Llama3 & 8b  & 98.96 & 85.52 & 37.08 & 60.22 & (-0.60) & & 99.52 & 95.24 & 60.82\\
    \midrule
    Llama3.1 & 8b  & 99.25 & 98.43 & 48.56 & 56.49 & (-3.36) & & 99.05 & 88.10 & 59.85\\
    \midrule
    Gemma & 7b  & 99.78 & 88.21 & 33.28 & 62.91 & (4.92) & & 99.52 & 89.52 & 57.99\\
    \midrule
    Starcoder-2-Instruct & 15b  & 98.36 & 92.84 & 57.14 & 64.40 & (-2.39) & & \textbf{100} & 99.05 & 66.79\\
    \midrule
    \multirow{3}{*}{DeepSeek-coder} & 1.3b  & 98.81 & 91.04 & 41.05 & 58.81 & (2.69) & & 94.76 & 90.0 & 56.12\\
    & 6.7b  & 94.78 & 92.99 & 45.09 & 65.60 & (3.81) & & 99.05 & 96.67 & 61.79\\
    & 33b  & 99.63 & 97.61 & 59.29 & 70.52 & (2.09) & & 100 & 99.52 & 68.43\\
    \midrule
    CodeQwen & 7b  & 94.78 & 92.99 & 61.05 & 65.60 & (3.81) & & 99.05 & 96.67 & 61.79 \\
    \bottomrule
  \end{tabular}
  }
\end{table*}

\subsection{Targeted Line and Branch Coverage}

Table~\ref{line-coverage} and Table~\ref{branch-coverage} show the evaluation results for the targeted line and branch coverage, respectively. For each subject LLM, we also include a baseline by excluding the information about the targeted lines/branches in the text prompt. For each program under test, we reuse the first test case generated for the overall coverage task and measure its coverage accuracy on the targeted lines/branches. The intuition is that, if an LLM could not outperform the baseline, it might be struggling with identifying the line/branch that is expected to cover when generating the test case.

Regarding the targeted line coverage, we find that the GPT-4 series has the best performance improvement (around 10\%) over their baselines. The best-performing LLM is GPT-4o, reaching a coverage accuracy of 80.97\% on average. 
We also find that six out of seventeen LLMs do not improve over their baseline and seven LLMs only have marginal improvement (less than 5\%). These results suggest that most LLMs may have trouble with multi-step reasoning. Specifically, to reach a specific line inside a branch, the LLM needs first to identify which branch the targeted line belongs to and then generate a valid test input to invoke this branch.


\begin{table*}[h]
  \caption{Results for targeted branch coverage. Results in parenthesis are the improvements over the baseline. We omit the correctness metrics of the baseline because they are the same as the targeted line coverage task.}
  \label{branch-coverage}
  \centering
  \scriptsize
  \setlength{\tabcolsep}{1.7pt}
  \scalebox{0.9}{
  \begin{tabular}{llcccccclclclclccccc}
    \toprule
    \multirow{3}{*}{Model} & \multirow{3}{*}{Size} & & \multicolumn{12}{c}{Targeted branch} & & \multicolumn{4}{c}{Baseline: no targeted branch} \\
    \cmidrule(lr){4-15} \cmidrule{17-20}
    & & & \multicolumn{3}{c}{Correctness} & & \multicolumn{8}{c}{Coverage} & & \multicolumn{4}{c}{Coverage} \\
    \cmidrule(lr){4-6} \cmidrule(lr){8-15} \cmidrule{17-20} 
    & & & syntax & execution & assertion & & \multicolumn{2}{c}{total} & \multicolumn{2}{c}{easy} & \multicolumn{2}{c}{medium} & \multicolumn{2}{c}{hard} & & total & easy & medium & hard \\
    \midrule
    GPT-3.5-turbo & N/A  & & \textbf{100} & 98.78 & 47.62 & & 70.40 & (4.38) & 82.93 & (0.40) & 65.33 & (1.77) & 50.77 & (14.23) & & 66.02 & 82.53 & 63.56 & \textbf{36.54}\\
    GPT-4 & N/A  &  & \textbf{100} & 98.17 & 61.88 & & 78.23 & (13.33) & 86.14 & (4.41) & 79.56 & (16.89) & 61.92 & (27.30) & & 64.90 & 81.73 & 62.67 & 34.62 \\
    GPT-4-turbo & N/A   & & \textbf{100} & 98.67 & 67.60 & & 80.77 & (15.15) & \textbf{88.35} & (5.42) & 79.11 & (16.00) & \textbf{67.69} & (33.07) & & 65.62 & 82.93 & 63.11 & 34.62\\
    GPT-4o & N/A & & \textbf{100} & \textbf{99.08} & \textbf{68.74} & & \textbf{80.87} & (12.61) & 87.55 & (3.21) & \textbf{83.11} & (11.55) & 66.15 & (31.53) & & \textbf{68.26} & \textbf{84.34} & \textbf{71.56} & 34.62\\
    GPT-4o-mini & N/A & & \textbf{100} & 99.39 & 57.65 & & 78.13 & (12.01) & 87.35 & (4.02) & 77.33 & (13.33) & 61.15 & (26.15) & & 66.12 & 83.33 & 64.00 & 35.00\\
    \midrule
    Gemini-1.0-pro & N/A & & \textbf{100} & 97.04 & 55.43 & & 68.97 & (5.80) & 82.13 & (2.21) & 69.78 & (6.22) & 43.08 & (12.31) & & 63.17 & 79.92 & 63.56 & 30.77\\
    \midrule
    \multirow{3}{*}{CodeLlama} & 7b & & \textbf{100} & 81.99 & 40.57 & & 50.97 & (-4.17) & 64.25 & (-8.04) & 51.11 & (-3.56) & 25.38 & (2.69) & & 55.14 & 72.29 & 54.67 & 22.69\\
    & 13b & & 99.29 & 82.91 & 52.86 & & 51.58 & (-4.68) & 64.86 & (-7.83) & 46.67 & (-11.55) & 30.39 & (5.78) & & 56.26 & 72.69 & 58.22 & 24.61\\
    & 34b & & 99.39 & 95.02 & 42.12 & & 63.17 & (5.69) & 78.51 & (2.20) & 60.44 & (6.22) & 36.15 & (11.92) & & 57.48 & 76.31 & 54.22 & 24.23\\
    \midrule
    Llama3 & 8b & & 98.88 & 84.94 & 37.93 & & 58.39 & (-0.31) & 73.09 & (-0.61) & 59.11 & (0.89) & 29.26 & (-1.12) & & 58.70 & 73.70 & 58.22 & 30.38\\
    \midrule
    Llama3.1 & 8b & & 99.49 & 85.86 & 48.20 & & 58.09 & (-0.71) & 69.08 & (-6.42) & 57.33 & (2.22) & 37.69 & (7.69) & & 58.80 & 75.50 & 55.11 & 30.00 \\
    \midrule
    Gemma & 7b & & 99.59 & 85.35 & 37.47 & & 56.15 & (1.11) & 71.89 & (2.01) & 49.78 & (0.45) & 31.54 & (0.00) & & 55.04 & 69.88 & 49.33 & 31.54\\
    \midrule
    Starcoder-2-Instruct & 15b  & & 98.68 & 95.42 & 64.63 & & 64.19 & (-0.41) & 78.71 & (0.20) & 63.56 & (-1.33) & 36.92 & (-0.77) & & 64.60 & 78.51 & 64.89 & 37.69\\
    \midrule
    \multirow{3}{*}{DeepSeek-coder} & 1.3b & & 97.05 & 89.32 & 41.11 & & 54.22 & (0.81) & 68.67 & (1.20) & 52.89 & (-1.78) & 27.69 & (2.31) & & 53.41 & 67.47 & 54.67 & 25.38\\
    & 6.7b & & 96.74 & 93.79 & 43.91 & & 66.43 & (7.22) & 77.11 & (5.62) & 69.33 & (4.89) & 43.46 & (12.31) & & 59.21 & 71.49 & 64.44 & 31.15\\
    & 33b & & \textbf{100} & 97.05 & 55.43 & & 68.46 & (2.54) & 80.12 & (-2.21) & 66.22 & (4.00) & 48.08 & (10.39) & & 65.92 & 82.33 & 62.22 & 37.69\\
    \midrule
    CodeQwen & 7b & & 99.49 & 95.02 & 61.76 & & 65.82 & (0.51) & 81.12 & (1.60) & 63.56 & (-3.55) & 38.46 & (1.92) & & 65.31 & 79.52 & 67.11 & \textbf{36.54}\\
    \bottomrule
  \end{tabular}
  }
\end{table*}

We observe a similar trend in the targeted branch coverage (Table~\ref{branch-coverage}). Specifically, the GPT-4 series has the best performance improvement (12\%\textasciitilde15\%) over their baselines. GPT-4o is the best-performing LLM, which can cover 80.87\% branches, respectively. By contrast, eight LLMs only exhibit marginal improvements and four LLMs do not improve compared with the baselines. Regarding branches with different difficulties, we find that branches more likely to be covered by random test cases are also easier for LLMs to cover (recall we use random testing to label each branch's difficulty level in \S~\ref{subsec:dataset}). The GPT-4 series shows the smallest performance gap between branches with different difficulty levels. We also notice that twelve out of sixteen LLMs show the largest performance improvements over the baselines on ``hard'' branches. These results indicate that providing target branch information can indeed help us to cover branches that are hard to reach by random testing.

\subsection{Targeted Path Coverage}
Table~\ref{path-coverage} presents the results of the targeted path coverage task. We adopt a similar baseline as in our targeted line/branch coverage tasks by excluding the targeted path in the text prompts. Overall, GPT-4o and Gemini-1.0-pro have the best performance on the path coverage, reaching 56.67\% and 56.09\% on average, respectively. However, they do not outperform their baselines. Generally, we do not find any LLMs that show obvious performance improvement (more than 5\%) on the path coverage compared with the baselines. Nine out of sixteen LLMs do not outperform the baselines. Regarding the path similarity, we also do not find any LLMs exhibiting large performance improvement compared with the baselines. These results suggest that comprehending the program logic and identifying a specific execution path is still a challenging task for the current LLMs.

Targeted path coverage is considerably more complicated compared with overall coverage and targeted line/branch coverage. Specifically, the LLM needs to identify a sequence of multiple branches, and create a test input that can execute these branches following a certain order, which is challenging even for human programmers. 

\begin{table*}[h]
  \caption{Results for targeted path coverage. Results in parenthesis are the improvements over the baseline.}
  \label{path-coverage}
  \centering
  \scriptsize
  \scalebox{0.9}{
  \begin{tabular}{llccccccccc}
    \toprule
    \multirow{2}{*}{Model} & \multirow{2}{*}{Size} & & \multicolumn{5}{c}{Given target path} & & \multicolumn{2}{c}{Baseline: no target path} \\
    \cmidrule{4-8} \cmidrule{10-11}
    & & & syntax & execution & assertion & path cov & path similarity & & path cov & path similarity\\
    \midrule
    GPT-3.5-turbo & N/A & & 99.88 & 98.95 & 49.58 & 49.30 (-5.97) & 77.35 (-2.39) & & 55.27 & 79.74\\
    GPT-4 & N/A & & \textbf{100} & 99.18 & 61.71 & 54.10 (-0.94) & 80.77 (3.23) & & 55.04 & 77.54\\
    GPT-4-turbo & N/A & & \textbf{100} & 99.41 & 63.00 & 50.47 (-3.74) & 79.82 (1.08) & & 54.21 & 78.74  \\
    GPT-4o & N/A & & \textbf{100} & 99.53 & \textbf{70.62} & \textbf{56.67} (-1.76) & \textbf{82.35} (1.29) & & 58.43 & \textbf{81.06}\\
    GPT-4o-mini & N/A & & \textbf{100} & \textbf{99.77} & 52.93 & 51.87 (-2.58) & 80.09 (0.67) & & 54.45 & 79.42 \\
    \midrule
    Gemini-1.0-pro & N/A & & \textbf{100} & 96.02 & 54.51 & 56.09 (0.70) & 77.59 (-0.23) & & 55.39 & 77.82\\
    \midrule
    \multirow{3}{*}{CodeLlama} & 7b & & 99.76 & 90.98 & 39.63 & 41.57 (-1.05) & 67.66 (0.14) & & 42.62 & 67.52\\
    & 13b & & 99.18 & 94.15 & 44.80 & 40.28 (-4.10) & 64.63 (-4.98) & & 44.38 & 69.61\\
    & 34b & & 98.95 & 96.25 & 41.11 & 48.01 (2.93) & 72.33 (2.69) & & 45.08 & 69.64\\ 
    \midrule
    Llama3 & 8b & & 98.24 & 89.46 & 33.72 & 41.92 (1.29) & 68.03 (0.40) & & 40.63 & 67.63\\
    \midrule
    Llama3.1 & 8b & & 99.88 & 95.78 & 39.64 & 44.02 (-1.41) & 72.51 \textbf{(4.24)} & & 45.43 & 68.27 \\
    \midrule
    Gemma & 7b & & \textbf{100} & 88.06 & 29.99 & 37.11 \textbf{(4.09)} & 64.54 (2.29) & & 33.02 & 62.25\\
    \midrule
    Starcoder-2-Instruct & 15b & & 96.83 & 90.28 & 47.86 & 48.48 (-5.38) & 70.91 (-6.78) & & 53.86 & 77.69\\
    \midrule
    \multirow{3}{*}{DeepSeek-coder} & 1.3b & & 97.89 & 88.99 & 40.27 & 40.16 (0.46) & 64.91 (0.67) & & 39.70 & 64.24\\
    & 6.7b & & 99.06 & 95.90 & 43.49 & 53.04 (0.23) & 76.77 (1.56) & & 52.81 & 75.21\\
    & 33b & & 100 & 96.49 & 63.28 & 54.10 (-4.33) & 77.99 (-2.73) & & 58.43 & 80.72\\
    \midrule
    CodeQwen & 7b & & 99.77 & 94.96 & 62.87 & 55.97 (-3.16) & 77.46 (-2.67) & & \textbf{59.13} & 80.13\\
    \bottomrule
  \end{tabular}
  }
\end{table*}

\subsection{Advanced Prompting}

Advanced prompting techniques, such as in-context learning \cite{brown2020language} and chain-of-thought (COT) \cite{wei2022chain}, can improve the performance of LLMs on language understanding and generation. We further conduct a study on the influence of different prompting strategies on \method. In this advanced prompt setting, we adopt an explicit two-step COT for the targeted line coverage task. LLMs are first asked to generate the conditions that need to be satisfied when the target line is executed. Then, we ask LLMs to generate a test case that satisfies these conditions. We provide a one-shot example of the reasoning process, which is created from the solution of a LeetCode easy-level problem (not included in the \method dataset). The complete prompt template for this setting is shown in Appendix~\ref{appendix:cot}.

\begin{table}[h]
  \caption{Results for two-step COT prompting on targeted line coverage. The results in parenthesis are the improvements over the basic prompting setting.}
  \label{cot-linecov}
  \centering
  \scriptsize
  \scalebox{0.9}{
  \begin{tabular}{llcccc}
    \toprule
    Model & Size &  syntax & execution & assertion & line coverage\\
    \midrule
    GPT-3.5-turbo & N/A  & 99.70 & 98.13 & 47.86 & 71.79 (4.03)\\
    GPT-4o & N/A  & 100 & 98.66 & 63.37 & 84.85 (3.88)\\
    GPT-4o-mini & N/A  & 100 & 97.84 & 49.73 & 76.72 (-0.22)\\
    \midrule 
    Llama3 & 8b  & 99.93 & 87.99 & 35.37 & 62.54 (2.32)\\
    \midrule
    Llama3.1 & 8b  & 99.70 & 97.24 & 38.14 & 64.48 (7.99) \\
    \midrule
    DeepSeek-coder & 6.7b  & 99.92 & 96.79 & 47.82 & 65.07 (-0.53)\\
    \bottomrule

  \end{tabular}
  }
\end{table}

Table~\ref{cot-linecov} shows the results of our COT prompting on the targeted line coverage task. Because the cost of COT is significantly higher than basic prompting, we only run experiments on several cost-efficient models, and omitted expensive proprietary models or large open-source models. For most models (except GPT-4o-mini and DeepSeek-coder 6.7b), COT can improve the performances on target line coverage. This suggests that building more complex LLM pipelines or agents for test case generation is worth investigating in the future.

\begin{figure}
    \includegraphics[width=\columnwidth]{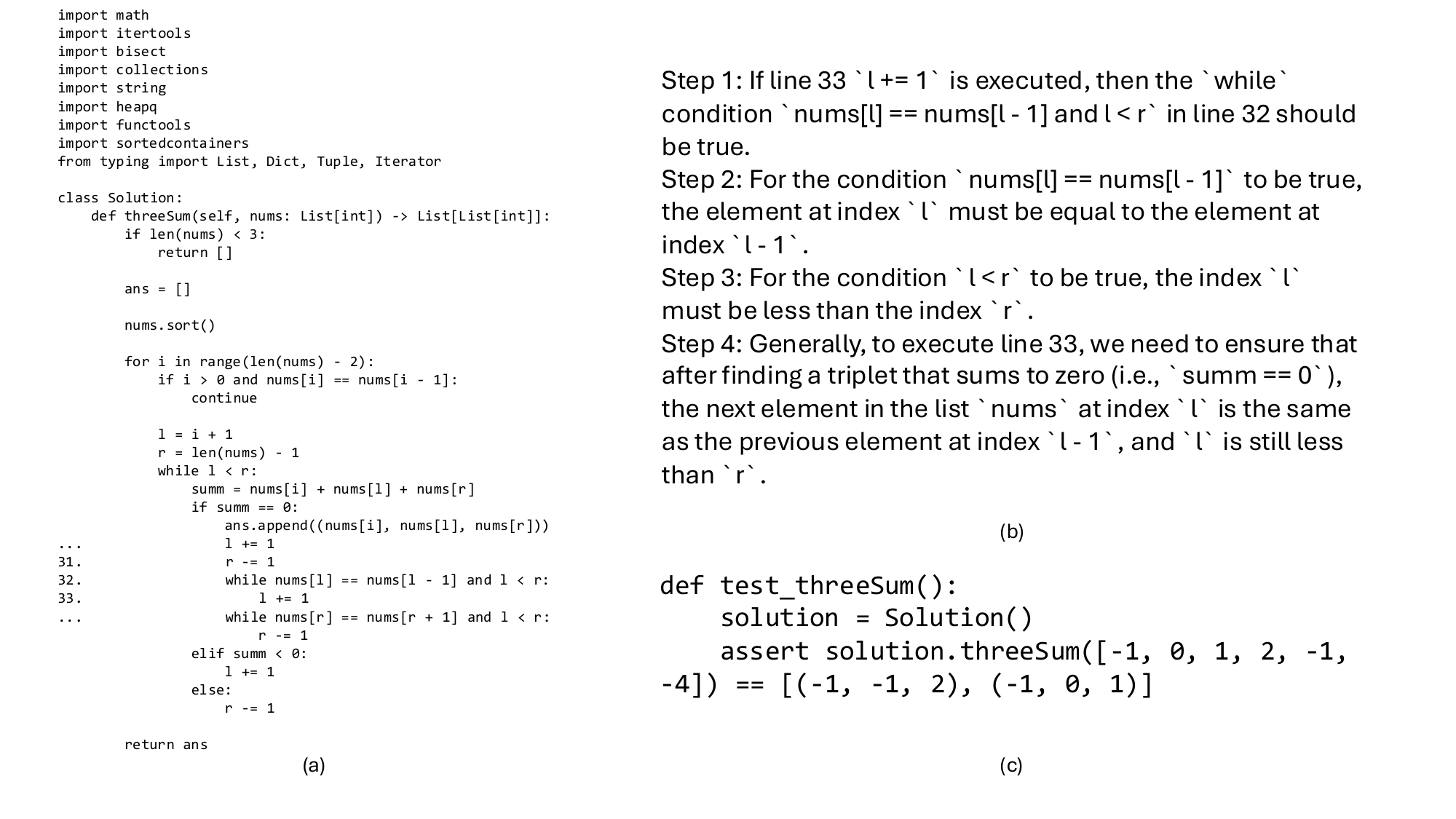}
  \caption{Example of a generated test case that failed to cover the target line. (a): the program under test. (b): LLM-generated reasoning steps. (c): LLM-generated test cases based on reasoning steps.} \label{fig:cover-error}
\end{figure}

With the two-step COT setting, we can have a detailed analysis of the reason behind failures in generated test cases. Figure~\ref{fig:cover-error} demonstrates a test case generated by GPT-4o that failed to cover the target line: line 33. We find that although the LLM is capable of generating correct conditions (Figure~\ref{fig:cover-error} (b)) for covering the target line, the generated test case did not satisfy those conditions, suggesting that the LLM's code generation ability needs further improvement. In this case, the generated test case (Figure~\ref{fig:cover-error} (c)) does not satisfy the condition `\texttt{nums[l] == nums[l - 1]}'.

\section{Related Work}

\textbf{Code-related Benchmarks for LLMs.} In recent years, researchers have endeavored to develop more rigorous and comprehensive evaluation frameworks for LLMs on coding abilities from various perspectives. One of the earliest attempts is HumanEval~\cite{chen2021evaluating}, which consists of 164 hand-craft programming challenges that evaluate LLMs' ability to understand natural language descriptions and generate the corresponding functional correct code.
Since then, there have been several studies attempting to construct benchmarks with more diverse problems~\cite{austin2021program}, more rigorous evaluations~\cite{liu2024your}, and more complex scenarios~\cite{lai2023ds, zheng2023codegeex, li-etal-2024-deveval}. Beyond these established code-generation scenarios, numerous studies are expanding their focus to include a broader range of real-world applications, such as reviewing code~\cite{li2022automating}, performing repo-level code completion~\cite{liu2023repobench,zhang-etal-2023-repocoder, guo2023longcoder, ding2024crosscodeeval}, and resolving GitHub issues~\cite{jimenez2023swe}. While all the aforementioned studies examine the coding abilities of LLMs from different perspectives, none specifically target test case generation, a crucial phase in the software engineering lifecycle. The most relevant study is DevBench~\cite{li2024devbench}, which evaluates LLMs across software development stages, including testing. Unlike DevBench, our benchmark provides more comprehensive evaluations specifically tailored to test case generation using coverage-guided tasks and includes a broader range of studied models.

\textbf{LLMs for Software Testing.} Recent studies have extensively utilized LLMs to develop efficient and effective testing pipelines for various software applications~\cite{xia2023universal, wang2024software}. Unit test case generation~\cite{schafer2023empirical}, which aims to test individual software components independently, is the primary focus of current LLM-aided software testing. One line of research tries to pre-train/fine-tune LLMs on focal methods and related assertion statements to enhance their test-generation capabilities~\cite{alagarsamy2023a3test, hashtroudi2023automated, rao2023cat, steenhoek2023reinforcement}. Although effective, these methods can be cost-intensive and challenging to scale.
Alternatively, some researchers focus on crafting effective prompts that instruct LLMs to analyze relevant information~\cite{yuan2023no, xie2023chatunitest, zhang2023well, li2023prompting, dakhel2024effective, ryan2024code, liu2024llm, pizzorno2024coverup, wang2024python} or documentation~\cite{vikram2023can, plein2024automatic}, or integrate LLMs with traditional software testing tools~\cite{lemieux2023codamosa}.

\section{Conclusion}
We present \method, a novel benchmark for evaluating automated test case generation with LLMs for Python programs. 
Based on this dataset, we propose three different tasks and standardized evaluation pipelines. Our targeted coverage tasks enable the assessment on the LLM's capabilities in comprehending complex program logic and execution path and generating test cases following the tester's intent, which is not considered in previous works on either code generation or test case generation with LLMs.

We further conduct extensive experiments with seventeen popular LLMs on {\method}. We find that although LLMs can achieve high overall coverage by generating diverse test cases, generating test cases to cover a specific element is still challenging. Our results reveal that there is a common lack of abilities in comprehending program logic among current LLMs, despite their promising performance on other code-related tasks (e.g., code generation). 



\section*{Limitations}

As a pioneering work of benchmarking LLM-based test case generation, our \method still has a few limitations. Here, we will discuss these limitations and how we addressed them in our work (or in the future).

First, the current \method is only limited to Python. Although the solutions in LeetCode are written in multiple languages, we find that their adopted algorithm and logical structures are largely the same. We believe the behaviors of LLMs in other languages of LeetCode solutions will be similar to those of Python, which we aim to verify in the future.

The second limitation is that our \method dataset is created from online programming problems, which may be different from real-world scenarios. We argue that at the current stage of LLM for test case generation, datasets from programming problems are still important. First, many LLMs in real-world test case generation still struggle with the correctness problem (whether the generated test case can be executed), which makes it too early to consider the coverage problem. For example, in \cite{yuan2023no}, ChatGPT only achieves 42.1\% success rate in compilation and 24.8\% in execution. In contrast, on \method, proprietary LLMs such as GPT-4 can achieve near 100\% accuracies in execution (although some open-source LLMs still have difficulties in generating correctly formatted test cases), which allows researchers to focus on how to improve test coverage. Second, compared to real-world software testing datasets, programs in \method have more complex control flow structures, which allow us to have a deeper study on how LLMs can reason about branches/loops in programs. For example, the real-world Python test case generation dataset CodaMOSA \cite{lemieux2023codamosa} has an average cyclomatic complexity of 5.85, while the average complexity of our \method dataset is 13.35.


\section*{Ethical Discussion}

Regarding the dataset, our dataset is built upon user-written solutions for LeetCode problems. These solutions are stored in a GitHub repository licensed with the MIT license, so we are granted permission to create our own dataset from this repository.

Regarding the use of automated systems, the automatic tools we used to create our dataset are all rule-based tools with no bias introduced.

The research of LLMs for test case generation may encourage the software development industry to use LLMs instead of human developers for software testing. However, our findings in the paper suggest that existing LLMs still encounter various difficulties in generating correct test cases with accurate target test coverage. As software testing in real-world practice may introduce new questions not discussed in this paper, thus the impact of this paper on the industry community is still limited and not likely to cause major concerns.

Also, using LLMs for automated software testing may raise security concerns. As our dataset only consists of self-contained, single-file programs, there are no security vulnerabilities in our dataset that can be exploited by LLM-generated test cases. However, if we extend the scope of our dataset to real-world software in the future, the security of experiments should be carefully considered.

\bibliography{custom}

\appendix


\clearpage
\section{Prompt Templates}
\label{appendix:template}
The prompt templates for \method tasks are shown as follows. For the targeted line/branch/path coverage tasks, we add line numbers to both the program under test and the target information in order to accurately locate the position of the target line/branch/path. Notice that our prompt template is only a primary setting without advanced prompting techniques such as few-shot examples or chain-of-thought reasoning, and we encourage future researchers to design more advanced prompts for \method.

\subsection{Prompt Template for Overall Coverage}

\begin{tcolorbox}[size=title,breakable]
Please write a test method for the function `\{func\_name\}' given the following program under test and function description. Your answer should only contain one test input.

Program under test:

----

\{program\}

----

Function description for `\{func\_name\}':

----

{description}

----

Your test method should begin with:

def test\_{func\_name}():
    
\ \ \ \ solution=Solution()
\tcblower
\textit{Prompt for generating the next test case:}

Generate another test method for the function under test. Your answer must be different from previously-generated test cases, and should cover different statements and branches.

\end{tcolorbox}

\subsection{Prompt Template for Targeted Line Coverage}

\begin{tcolorbox}[size=title,breakable]
Please write a test method for the function `\{func\_name\}' given the following program under test and function description. Your answer should only contain one test input.

Program under test:

----

\{program\}

----

Function description for `\{func\_name\}':

----

{description}

----

Your test case must cover line \{target\_line\}.

Your test method should begin with:

def test\_{func\_name}():
    
\ \ \ \ solution=Solution()

\end{tcolorbox}

\subsection{Prompt Template for Targeted Branch Coverage}
\begin{tcolorbox}[size=title,breakable]
Please write a test method for the function `\{func\_name\}' given the following program under test and function description. Your answer should only contain one test input.

Program under test:

----

\{program\}

----

Function description for `\{func\_name\}':

----

{description}

----

Your test case must cover the branch \{target\_branch\}.

Your test method should begin with:

def test\_{func\_name}():
    
\ \ \ \ solution=Solution()

\end{tcolorbox}

\subsection{Prompt Template for Targeted Path Coverage}

\begin{tcolorbox}[size=title,breakable]
Please write a test method for the function `\{func\_name\}' given the following program under test and function description. Your answer should only contain one test input.

Program under test:

----

\{program\}

----

Function description for `\{func\_name\}':

----

{description}
----

Your test case must cover the following execution path in function \{func\_name\}. The path is a sequence of branch conditions. When executing your test case, each branch condition in the target execution path must be satisfied sequentially.

Target execution path: \{target\_path\}

----

Your test method should begin with:

def test\_{func\_name}():
    
\ \ \ \ solution=Solution()

\end{tcolorbox}

\subsection{Prompt Template for Two-step COT}
\label{appendix:cot}

\begin{tcolorbox}[float*=b,title=Prompt template for generating conditions in the two-step COT,size=title,breakable,width=\textwidth]
Given a Python code snippet and a target line number, you are asked to generate reasoning steps to satisfy a specific line to be executed.

[Example]

Given the following code snippet:

\verb|```|Python

class Solution: \#1

\ \ \ \ def twoSum(self, nums: List[int], target: int) -> List[int]:    \#2
    
\ \ \ \ \ \ \ \ numMap = {{}} \#3
        
\ \ \ \ \ \ \ \ n = len(nums)   \#4
        
\ \ \ \ \#5
    
\ \ \ \ \ \ \ \ for i in range(n):  \#6
        
\ \ \ \ \ \ \ \ \ \ \ \ numMap[nums[i]] = i \#7
            
\ \ \ \ \#8
    
\ \ \ \ \ \ \ \ for i in range(n):  \#9
        
\ \ \ \ \ \ \ \ \ \ \ \ complement = target - nums[i]   \#10
            
\ \ \ \ \ \ \ \ \ \ \ \ if complement in numMap and numMap[complement] != i:    \#11
            
\ \ \ \ \ \ \ \ \ \ \ \ \ \ \ \ return [i, numMap[complement]]  \#12
                
\ \ \ \ \#13
    
\ \ \ \ \ \ \ \ return []   \#14
        
\verb|```|

Identify when executing funtion twoSum, what conditions need to be satisfied if line 12 is to be executed.

Answer:

<cond>

Step 1: If line 12 `return [i, numMap[complement]]` is executed, then the `if` condition `(complement in numMap and numMap[complement] != i)` in line 11 shoud be true.

Step 2: If condition `complement in numMap` is true, at least one `target - nums[i]` in line 10 equals an element in nums, which means there exists two elements in `nums` that their sum is equal to `target`.

Step 3: If condition `numMap[complement] != i` is ture, then `numMap[target - nums[i]] != i`, meaning that the index of `target - nums[i]` is not equal to `i`.

Step 4: Generally, to execute line 12, we need to ensure that there exists two different elements in `nums` that their sum is equal to `target`.

<\textbackslash cond>

[\textbackslash Example]

In a similar fashion, identify the conditions that need to be satisfied when line {targetline} is to be executed for the following Python code.

\verb|```|Python

\{program\}

\verb|```|

Surround your answer with <cond> and <\textbackslash cond>.
\end{tcolorbox}

\begin{tcolorbox}[float*=b,title=Prompt template for generating test case in the two-step COT,size=title,breakable,width=\textwidth]
For the given code snippet and a list of conditions need to be satisfied, generate a test case that will satisfiy these conditions. Here is an example:

[Example]

Code:

\verb|```|Python

class Solution: \#1

\ \ \ \ def twoSum(self, nums: List[int], target: int) -> List[int]:    \#2
    
\ \ \ \ \ \ \ \ numMap = {{}} \#3
        
\ \ \ \ \ \ \ \ n = len(nums)   \#4
        
\ \ \ \ \#5
    
\ \ \ \ \ \ \ \ for i in range(n):  \#6
        
\ \ \ \ \ \ \ \ \ \ \ \ numMap[nums[i]] = i \#7
            
\ \ \ \ \#8
    
\ \ \ \ \ \ \ \ for i in range(n):  \#9
        
\ \ \ \ \ \ \ \ \ \ \ \ complement = target - nums[i]   \#10
            
\ \ \ \ \ \ \ \ \ \ \ \ if complement in numMap and numMap[complement] != i:    \#11
            
\ \ \ \ \ \ \ \ \ \ \ \ \ \ \ \ return [i, numMap[complement]]  \#12
                
\ \ \ \ \#13
    
\ \ \ \ \ \ \ \ return []   \#14

\verb|```|

Conditions:

Step 1: If line 12 `return [i, numMap[complement]]` is executed, then the `if` condition `(complement in numMap and numMap[complement] != i)` in line 11 shoud be true.

Step 2: If condition `complement in numMap` is true, at least one `target - nums[i]` in line 10 equals an element in nums, which means there exists two elements in `nums` that their sum is equal to `target`.

Step 3: If condition `numMap[complement] != i` is ture, then `numMap[target - nums[i]] != i`, meaning that the index of `target - nums[i]` is not equal to `i`.

Step 4: Generally, to execute line 12, we need to ensure that there exists two different elements in `nums` that their sum is equal to `target`.

Generated test case:

\verb|```|Python

def test\_twoSum():

\ \ \ \ solution = Solution()
    
\ \ \ \ assert solution.twoSum([2,7,11,15], 9) == [0, 1]
    
\verb|```|

[\textbackslash Example]

In a similar fashion, generate a test case for the following code snippet and conditions. Your test function should be named `test\_{func\_name}`.
Code:

\verb|```|Python

\{program\}

\verb|```|

Conditions:

\{conditions\}

You should only generate the test case, without any additional explanation.
\end{tcolorbox}

\clearpage
\section{Targeted Line/Branch Identification}
\label{appendix:branch}
The complete algorithm for extracting targeted lines/branches from a program under test is shown in Algorithm 2. At a high level, we first extract all conditional branches by locating the branches starting with conditional operators (i.e., `\texttt{if}', `\texttt{elif}', and `\texttt{else}') through parsing the program's abstract syntax tree. For each branch, we record the line numbers of its first and last lines (e.g., Lines 1:5) as one targeted branch. Then, we record the line numbers of all lines (except the line that only includes the `\texttt{else}' operator) within this branch as the targeted lines (e.g., [1, 2, 3, 4, 5]). We repeat this process until finishing parsing all branches of a program. 

\begin{algorithm}[h]
\label{alg:branch_identify}
        \SetAlgoLined 
        \caption{Targeted Line/Branch Identification.}
        \KwIn{Program with $L$ lines: $p=\left\{s_{1}, s_{2}, ..., s_{L}\right\}$}
        \KwOut{Target lines $ls$, target branches $bs$}
        $ls$ = [], $bs$ = [], $i=1$\;
        \While {$i<=L$}{
            \If{$s_{i}$ starts with `\texttt{if}', `\texttt{elif}', or `\texttt{else}'}{
                $curent\_branch$ = [] \;
                $j=i$ \;
                \Repeat{$s_{j}$ {\bf not} in this branch}{
                    $curent\_branch$.append($j$) \;
                    $j=j+1$
                }
                $bs$.append($curent\_branch$)\;
            }            
        }
        \For{$target\_branch$ in $bs$}{
            \For{line $s_{i}$ in $target\_branch$}{
                \If{$s_{i}$ is inside a branch {\bf and} {\bf not} $s_{i}$ starts with '\texttt{else}'}{
                    $ls$.append($i$)\;
                }
            }
        }
        \Return{$ls$, $bs$}
    \end{algorithm}

\section{Error Analysis}

\label{sec:error-exec}
\begin{figure}
  \centering
    \includegraphics[width=\columnwidth]{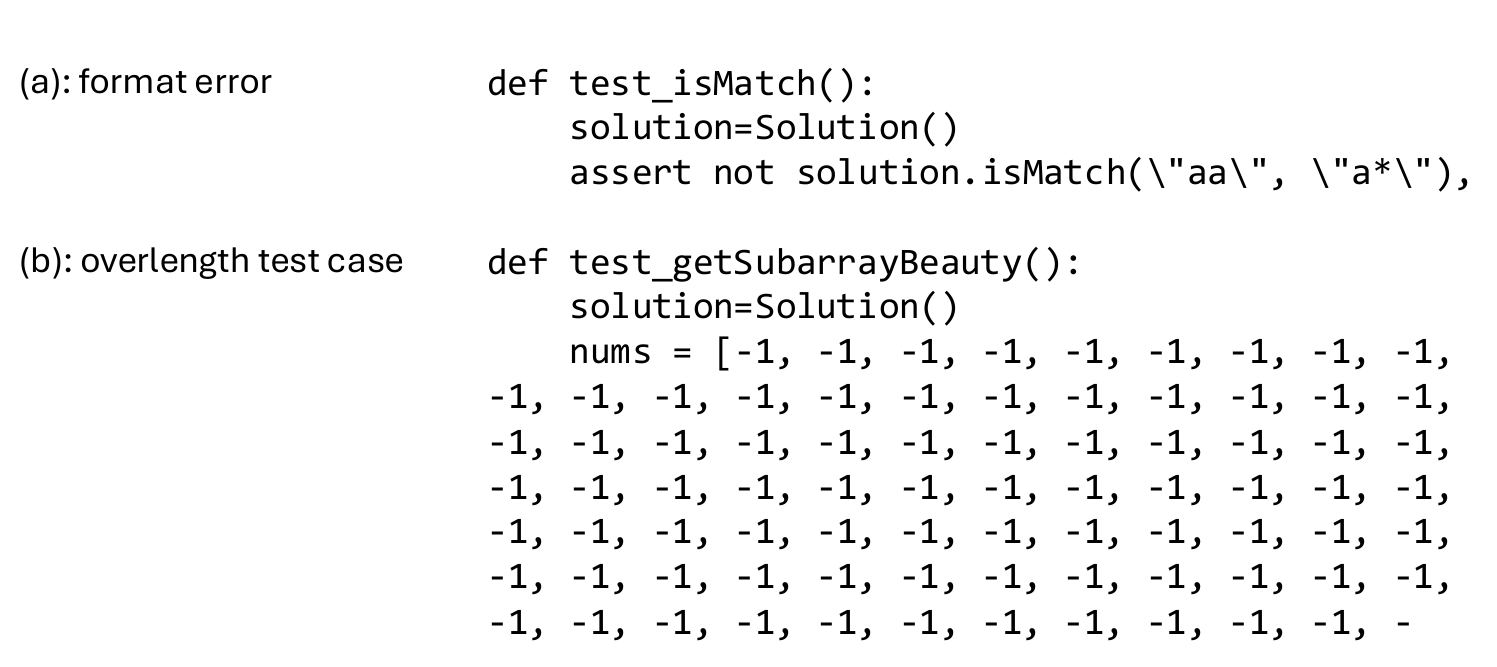}
  \caption{Examples of erroneous test cases generated by LLMs.} \label{fig:exec-error}
\end{figure}

For the failure of LLMs in generating test cases that failed to execute, we choose Llama3 as the example. Figure~\ref{fig:exec-error} in our attached pdf file shows several examples of failed test cases generated by Llama 3. Figure~\ref{fig:exec-error} (a) shows an example with a slight syntax error: it generated a redundant comma at the end of the last statement. Figure~\ref{fig:exec-error} (b) is another common type of error: the LLM generates an endless statement by repeating a simple pattern. In our post-processing statements, we remove the last statement if it is uncompilable. These erroneous statements are removed and result in empty test cases, which are counted as execution errors. We find that all execution errors in Llama 3-8b for targeted line coverage are made up of these two types of errors.

\section{Data Leakage Analysis}
\label{sec:data-leakage}

We choose GPT-4o as an example to study the potential of data leakage. The training data of GPT-4o covers up to October 2023, so we filter the problems from our dataset released after Oct 2023, which results in a total of 21 problems. Correspondingly, we also create a subset with 21 oldest problems which are released before Oct 2023. 

For the problems released after Oct 2023, in their 49 official test cases, we found none of them appeared in the generated test cases. On the contrary, for the 21 problems before Oct 2023, 35 out of 52 official test cases have been found in the generated test cases. However, as the LLM has generated 20 different test cases for each problem (which means 420 test cases for 21 problems), the issue of copying official test cases is minor.
We further measure the overall coverage for all problems before/after Oct 2023, the results are shown in Table~\ref{tab:data-leak}.

\begin{table}[htbp]
  \caption{Coverage metrics of the overall coverage task with data source before and after Oct 2023.}
  \label{tab:data-leak}
  \centering
  \scriptsize
  \scalebox{1.0}{
  \begin{tabular}{lcccc}
    \toprule
    \multirow{2}{*}{Model} & \multicolumn{2}{c}{Before Oct 2023} & \multicolumn{2}{c}{After Oct 2023} \\
    & line & branch & line & branch\\
    \midrule
    GPT-4o & 98.74 & 97.24 & 97.79 & 96.38\\
    \bottomrule
  \end{tabular}
  }
\end{table}

We can see that the coverage metrics before/after Oct 2023 are similar, indicating that potential data leakage is not a major concern of TestEval.


\end{document}